\documentclass[preprint,12pt]{elsarticle}

\usepackage{amssymb}

\usepackage{amsthm}

\usepackage{amsmath}
\DeclareMathOperator{\tr}{tr}

\DeclareMathOperator{\supp}{supp}
\newcommand{\Slash}[1]{{\ooalign{\hfil/\hfil\crcr$#1$}}}
\numberwithin{equation}{section}

\usepackage{url}

\begin{document}

\begin{frontmatter}



\title{%
Ferrara--Zumino supermultiplet and the energy-momentum tensor in the lattice
formulation of 4D $\mathcal{N}=1$ SYM}


\author{Hiroshi Suzuki}
\ead{hsuzuki@riken.jp}
\address{%
%Quantum Hadron Physics Laboratory,
Theoretical Research Division, RIKEN Nishina Center, Wako 2-1, Saitama
351-0198, Japan}

\begin{abstract}
It is well-known that Noether currents in the classical four-dimensional
$\mathcal{N}=1$ supersymmetric Yang--Mills theory (4D $\mathcal{N}=1$ SYM),
i.e., the $U(1)_A$ current, the supersymmetry (SUSY) current and the
energy-momentum tensor, form a multiplet under SUSY, called the
Ferrara--Zumino supermultiplet. Inspired by this structure, we define the
energy-momentum tensor in the lattice formulation of 4D $\mathcal{N}=1$ SYM by
a renormalized super transformation of a lattice SUSY current. By using a
renormalized SUSY Ward--Takahashi relation, the energy-momentum tensor so
constructed is shown to be conserved in the quantum continuum limit. Our
construction of the energy-momentum tensor is very explicit and usable in
non-perturbative numerical simulations.
\end{abstract}
% Report number: RIKEN-QHP-47

\end{frontmatter}

% \linenumbers

\section{Introduction}
%\label{}
\label{sec:1}
Ideally, a non-perturbative formulation of a field theory with some symmetries
should provide, not only the definition of correlation functions, but also the
definition of renormalized Noether currents that generate correctly-normalized
symmetry transformations on renormalized fields. This is expressed by
renormalized Ward--Takahashi (WT) relations and, when these relations hold, one
may say that the symmetries are \emph{really\/} realized in quantum field
theory.

Quite often, however, the regularization procedure breaks the preferred
symmetries and, for this reason, it is generally very difficult to conclude
even the existence of such renormalized Noether currents (especially when the
symmetry transformations are non-linear). Even if such Noether currents are
assumed to exist, the explicit construction can be very cumbersome. In the
lattice formulation of supersymmetric theories, one encounters such a
situation, because the infinitesimal translation (that is a part of the SUSY
algebra) is broken by the spacetime lattice, implying also the breaking of
SUSY. Here, almost all fundamental symmetries that define the theory are
broken by the regularization.

Having the above general remark in mind, in the present paper, we study the
construction of Noether currents in the lattice formulation of the
four-dimensional $\mathcal{N}=1$ supersymmetric Yang--Mills theory (4D
$\mathcal{N}=1$ SYM)~\cite{Curci:1986sm} (see also Ref.~\cite{Kaplan:1983sk}
for an earlier consideration and Ref.~\cite{Montvay:2001aj} for a very readable
review). The classical Euclidean action of 4D $\mathcal{N}=1$ SYM is given
by\footnote{We basically follow the notation of~Ref.~\cite{Suzuki:2012pc}:
The sum over repeated indices is understood. Vector indices $\mu$, $\nu$,
\dots, run over $0$, $1$, $2$, $3$. $\epsilon_{\mu\nu\rho\sigma}$~denotes the
totally anti-symmetric tensor and~$\epsilon_{0123}=-1$. All gamma matrices are
hermitian and obey $\{\gamma_\mu,\gamma_\nu\}=2\delta_{\mu\nu}$. We define
$\gamma_5\equiv-\gamma_0\gamma_1\gamma_2\gamma_3$
and~$\sigma_{\mu\nu}\equiv[\gamma_\mu,\gamma_\nu]/2$. The charge conjugation
matrix~$C$ satisfies, $C^{-1}\gamma_\mu C=-\gamma_\mu^T$,
$C^{-1}\sigma_{\mu\nu}C=-\sigma_{\mu\nu}^T$, $C^{-1}\gamma_5C=\gamma_5^T$
and~$C^T=-C$. The generator of the gauge group $SU(N_c)$, $T^a$, is normalized
as $\tr(T^aT^b)=(1/2)\delta^{ab}$. $F_{\mu\nu}(x)$ is the field strength
$F_{\mu\nu}(x)\equiv\partial_\mu A_\nu(x)-\partial_\nu A_\mu(x)%
+ig[A_\mu(x),A_\nu(x)]$, where $g$~is the bare gauge coupling constant, and
$\Slash{D}\equiv\gamma_\mu D_\mu$, where
$D_\mu\equiv\partial_\mu+ig[A_\mu(x),]$ is the covariant derivative in the
adjoint representation. For the lattice theory, $x$, $y$, $z$, \dots\ denote
lattice points and $a$~is the lattice spacing; $\Hat{\mu}$ is the unit vector
in the $\mu$-direction. The link variable~$U_\mu(x)$ and the gauge
potential~$A_\mu(x)$ are related as
\begin{equation}
   U_\mu(x)=e^{iagA_\mu(x)}.
\label{eq:(1.1)}
\end{equation}
The forward and backward difference operators respectively are defined by
\begin{equation}
   \partial_\mu f(x)
   \equiv\frac{1}{a}\left[f(x+a\Hat{\mu})-f(x)\right],\qquad
   \partial_\mu^*f(x)
   \equiv\frac{1}{a}\left[f(x)-f(x-a\Hat{\mu})\right],
\label{eq:(1.2)}
\end{equation}
and the symmetric difference operator $\partial_\mu^S$ is defined by
\begin{equation}
   \partial_\mu^S\equiv\frac{1}{2}(\partial_\mu+\partial_\mu^*).
\label{eq:(1.3)}
\end{equation}}
\begin{equation}
   S_{\text{classical}}
   =\int d^4x\,\left\{
   \frac{1}{2}\tr\left[F_{\mu\nu}(x)F_{\mu\nu}(x)\right]
   +\tr\left[\Bar{\psi}(x)\Slash{D}\psi(x)\right]\right\},
\label{eq:(1.4)}
\end{equation}
where the adjoint fermion (gluino)~$\psi(x)$ on the Euclidean space is subject
of the constraint
\begin{equation}
   \Bar{\psi}(x)=\psi^T(x)(-C^{-1}),
\label{eq:(1.5)}
\end{equation}
to express the degrees of freedom of a Majorana fermion in the 4D Minkowski
space. The global SUSY~$\Bar{\delta}_\xi$ in the classical 4D $\mathcal{N}=1$
SYM is
\begin{align}
   \Bar{\delta}_\xi A_\mu(x)&=\Bar{\xi}\gamma_\mu\psi(x),&&
\notag\\
   \Bar{\delta}_\xi\psi(x)&=-\frac{1}{2}\sigma_{\mu\nu}\xi F_{\mu\nu}(x),&
   \Bar{\delta}_\xi\Bar{\psi}(x)
   &=\frac{1}{2}\Bar{\xi}\sigma_{\mu\nu}F_{\mu\nu}(x),
\label{eq:(1.6)}
\end{align}
where the Grassmann-odd constant spinor~$\xi$ obeys $\Bar{\xi}=\xi^T(-C^{-1})$.
Without containing the scalar field, 4D $\mathcal{N}=1$ SYM is the simplest
supersymmetric theory in four dimensions and we should understand the issue
raised at the beginning of this paper first in this example, before tackling
more complicated supersymmetric theories.

The most interesting result we will obtain below is a very explicit form of an
energy-momentum tensor on the lattice $\mathcal{T}_{\mu\nu}(x)$, given
by~Eqs.~\eqref{eq:(3.3)} and~\eqref{eq:(3.4)}, which is \emph{conserved\/} in
the quantum continuum limit. Our way of construction of this energy-momentum
tensor was suggested by the fact that in the classical 4D $\mathcal{N}=1$ SYM,
the Noether currents associated with the $U(1)_A$ symmetry, SUSY and
translational invariance, form a multiplet under SUSY, the Ferrara--Zumino (FZ)
supermultiplet~\cite{Ferrara:1974pz}.\footnote{See Chapter~20
of~Ref.~\cite{West:1990tg} for a very readable exposition. An interesting
application of the notion of the FZ multiplet in phenomenology was recently
considered in~Ref.~\cite{Kitano:2011fk}.} For the classical
theory~\eqref{eq:(1.4)}, we define the $U(1)_A$
current~$\Breve{\jmath}_{5\mu}(x)$, the SUSY current~$\Breve{S}_\mu(x)$ and the
energy-momentum tensor~$\Breve{T}_{\mu\nu}(x)$ by
\begin{align}
   \Breve{\jmath}_{5\mu}(x)
   &\equiv\tr\left[\Bar{\psi}(x)\gamma_\mu\gamma_5\psi(x)\right],
\notag\\
   \Breve{S}_\mu(x)
   &\equiv-\sigma_{\rho\sigma}\gamma_\mu\tr\left[\psi(x)F_{\rho\sigma}(x)\right],
\notag\\
   \Breve{T}_{\mu\nu}(x)
   &\equiv
   2\tr\left[F_{\mu\rho}(x)F_{\nu\rho}(x)\right]
   -\frac{1}{2}\delta_{\mu\nu}\tr\left[F_{\rho\sigma}(x)F_{\rho\sigma}(x)\right]
\notag\\
   &\qquad{}
   +\frac{1}{4}\tr\left[
   \Bar{\psi}(x)\left(\gamma_\mu\overleftrightarrow{D}_\nu
   +\gamma_\nu\overleftrightarrow{D}_\mu\right)\psi(x)
   \right]
   -\frac{1}{2}\delta_{\mu\nu}\tr\left[
   \Bar{\psi}(x)\overleftrightarrow{\Slash{D}}\psi(x)
   \right],
\label{eq:(1.7)}
\end{align}
where the definitions of~$\Breve{\jmath}_{5\mu}(x)$ and~$\Breve{S}_\mu(x)$ are
standard while, as well-known, the definition of the energy-momentum tensor is
large extent arbitrary. In Eq.~\eqref{eq:(1.7)}, we defined the energy-momentum
tensor by first introducing a background gravitational field
into the continuum action~\eqref{eq:(1.4)} and then taking a flat-space limit
after differentiating the action with respect to the gravitational field. Then
one finds, under the super transformation~\eqref{eq:(1.6)},
\begin{equation}
   \Bar{\delta}_\xi\Breve{\jmath}_{5\mu}(x)=\Bar{\xi}\gamma_5\Breve{S}_\mu(x),
\label{eq:(1.8)}
\end{equation}
and
\begin{align}
   \Bar{\delta}_\xi\Breve{S}_\mu(x)
   &=2\gamma_\nu\xi
   \biggl\{\Breve{T}_{\mu\nu}(x)
   +\frac{3}{4}\delta_{\mu\nu}
   \tr\left[\Bar{\psi}(x)\Slash{D}\psi(x)\right]
\notag\\
   &\qquad\qquad{}
   +\frac{1}{4}\epsilon_{\mu\nu\rho\sigma}
   \partial_\rho\Breve{\jmath}_{5\sigma}(x)
   +\frac{1}{4}\tr\left[\Bar{\psi}(x)\sigma_{\mu\nu}\Slash{D}\psi(x)\right]
   \biggr\}
\notag\\
   &\qquad{}
   +\gamma_5\gamma_\nu\xi\partial_\nu\Breve{\jmath}_{5\mu}(x)
\notag\\
   &\qquad{}
   +\frac{3}{2}\xi\tr\left[\Bar{\psi}(x)\gamma_\mu\Slash{D}\psi(x)\right]
   +\frac{3}{2}\gamma_5\xi\tr\left[\Bar{\psi}(x)\gamma_\mu\gamma_5
   \Slash{D}\psi(x)\right]
\notag\\
   &\qquad{}
   -\frac{1}{2}\gamma_5\gamma_\nu\xi\tr\left[\Bar{\psi}(x)
   \gamma_\mu\gamma_5\gamma_\nu\Slash{D}\psi(x)\right]
\notag\\
   &\qquad{}
   -\frac{1}{4}\sigma_{\nu\rho}\xi\tr\left[\Bar{\psi}(x)
   \gamma_\mu\sigma_{\nu\rho}\Slash{D}\psi(x)\right].
\label{eq:(1.9)}
\end{align}
Similarly, $\Bar{\delta}_\xi\Breve{T}_{\mu\nu}(x)$ becomes a linear combination
of~$\partial_\rho\Breve{S}_\sigma(x)$, up to terms being proportional to the
equation of motion of the (massless) gluino, $\Slash{D}\psi(x)$. By integrating
over the spatial coordinates of the $\mu=0$ component of the above relations,
one sees that they are consistent with the defining algebra between the SUSY
charge, the $U(1)_A$ charge (this is the $R$-charge) and the momentum, up to
the equation of motion $\Slash{D}\psi(x)=0$. Eqs.~\eqref{eq:(1.8)}
and~\eqref{eq:(1.9)} thus may be regarded as a basic characterization of
fundamental Noether currents in~Eq.~\eqref{eq:(1.7)}. By going one step
further, one may regard Eqs.~\eqref{eq:(1.8)} and~\eqref{eq:(1.9)} as defining
relations of the Noether currents. This is the approach we adopt in the present
paper. That is, we define a lattice energy-momentum
tensor~$\mathcal{T}_{\mu\nu}(x)$ by a (renormalized modified) super
transformation of a (renormalized) lattice SUSY current~$\mathcal{S}_\mu(x)$
that is conserved in the continuum limit under an appropriate gluino mass
tuning~\cite{Curci:1986sm}. Then it can be shown that the lattice
energy-momentum tensor so constructed is conserved in the continuum limit. This
conservation of the energy-momentum tensor is not quite trivial, because SUSY
is not a manifest symmetry with the lattice regularization. For the
conservation of~$\mathcal{T}_{\mu\nu}(x)$, a ``finiteness'' of a lattice super
transformation induced by~$\mathcal{S}_\mu(x)$ turns to be crucial. This
finiteness might be regarded as a necessary condition for the existence of a
renormalized 4D $\mathcal{N}=1$ SYM.

As the structure of our energy-momentum tensor is very explicit as
Eqs.~\eqref{eq:(3.3)} and~\eqref{eq:(3.4)} show, it is usable in actual
Monte Carlo simulations of 4D $\mathcal{N}=1$ SYM~\cite{Montvay:1996pz,%
Montvay:1997ak,Koutsoumbas:1997de,Kirchner:1998nk,Kirchner:1998mp,%
Campos:1999du,Feo:1999hw,Feo:1999hx,Farchioni:2000mp,Farchioni:2000kb,%
Farchioni:2001yn,Farchioni:2001yr,Farchioni:2001wx,Peetz:2002sr,%
Farchioni:2004fy,Demmouche:2008ms,Demmouche:2009ki,Demmouche:2010sf,%
Bergner:2011wf,Bergner:2012at,Fleming:2000fa,Giedt:2008xm,Endres:2009yp,%
Endres:2009pu,Kim:2011fw}. Physical quantities such as the viscosity (see, for
example, Ref.~\cite{Sakai:2007cm} and references therein) might be measured
from the two-point function of our lattice energy-momentum tensor.

Our argument below is quite general and rather independent of the details of
a lattice formulation. We assume the locality of the lattice action, that
consists of the sum of a gauge boson (gluon) action~$S_{\text{gluon}}$,
the kinetic term~$S_{\text{gluino}}$ and the mass term~$S_{\text{mass}}$ of the
gauge fermion (gluino),\footnote{Although we use the language of a
four-dimensional lattice theory in the present paper, we believe that our
argument can be transcribed for a
domain-wall-type~\cite{Kaplan:1992bt,Shamir:1993zy} five-dimensional
setting~\cite{Nishimura:1997vg,Maru:1997kh,Kaplan:1999jn} without much
difficulty.}
\begin{equation}
   S\equiv S_{\text{gluon}}+S_{\text{gluino}}+S_{\text{mass}},
\label{eq:(1.10)}
\end{equation}
where
\begin{equation}
   S_{\text{gluino}}\equiv
   a^4\sum_x\tr\left[\Bar{\psi}(x)D\psi(x)\right],
\label{eq:(1.11)}
\end{equation}
where $D$ is a lattice Dirac operator ($(C^{-1}D)^T=-C^{-1}D$), and
\begin{equation}
   S_{\text{mass}}\equiv
   a^4\sum_xM\tr\left[\Bar{\psi}(x)\psi(x)\right].
\label{eq:(1.12)}
\end{equation}
We further assume that the lattice action~\eqref{eq:(1.10)} is invariant under
the hypercubic group~$H(4)$, including the parity transformation~$\mathcal{P}$
defined by
\begin{align}
   U_0(x_0,\Vec{x})&\xrightarrow{\mathcal{P}}U_0(x_0,-\Vec{x}),&
   U_k(x_0,\Vec{x})&\xrightarrow{\mathcal{P}}U_k^\dagger(x_0,-\Vec{x}-a\hat k),
\notag\\
   \psi(x_0,\Vec x)&\xrightarrow{\mathcal{P}}i\gamma_0\psi(x_0,-\Vec x),&
   \Bar{\psi}(x_0,\Vec x)&\xrightarrow{\mathcal{P}}
   -i\Bar{\psi}(x_0,-\Vec x)\gamma_0,
\label{eq:(1.13)}
\end{align}
where and in what follows, $\Vec{x}\equiv(x_1,x_2,x_3)$ and~$k$, $l=1$ or~$2$
or~$3$.

In the following discussion, we make frequent use of a lattice transcription of
the field strength,
\begin{equation}
   \left[F_{\mu\nu}\right]^L(x)\equiv2\tr\left[P_{\mu\nu}(x)T^a\right]T^a,
\label{eq:(1.14)}
\end{equation}
that is defined from the clover plaquette $P_{\mu\nu}(x)$,
\begin{equation}
   P_{\mu\nu}(x)\equiv\frac{1}{4}\sum_{i=1}^4\frac{1}{2ia^2g}
   \left[U_{i\mu\nu}(x)-U_{i\mu\nu}^\dagger(x)\right],
\label{eq:(1.15)}
\end{equation}
where
\begin{align}
   U_{1\mu\nu}(x)
   &\equiv U_\mu(x)U_\nu(x+a\hat\mu)U_\mu^\dagger(x+a\hat\nu)U_\nu^\dagger(x),
\notag\\
   U_{2\mu\nu}(x)
   &\equiv U_\nu(x)U_\mu^\dagger(x-a\hat\mu+a\hat\nu)
   U_\nu^\dagger(x-a\hat\mu)U_\mu(x-a\hat\mu),
\notag\\
   U_{3\mu\nu}(x)
   &\equiv U_\mu^\dagger(x-a\hat\mu)U_\nu^\dagger(x-a\hat\mu-a\hat\nu)
   U_\mu(x-a\hat\mu-a\hat\nu)U_\nu(x-a\hat\nu),
\notag\\
   U_{4\mu\nu}(x)
   &\equiv U_\nu^\dagger(x-a\hat\nu)U_\mu(x-a\hat\nu)
   U_\nu(x+a\hat\mu-a\hat\nu)U_\mu^\dagger(x).
\label{eq:(1.16)}
\end{align}
Note that $[F_{\mu\nu}]^L(x)$ is traceless by construction~\eqref{eq:(1.14)}.
Under the parity, $[F_{\mu\nu}]^L(x)$ transforms in an identical manner as the
continuum field strength. That is, we have
\begin{equation}
   \left[F_{0k}\right]^L(x_0,\Vec{x})
   \xrightarrow{\mathcal{P}}-\left[F_{0k}\right]^L(x_0,-\Vec{x}),\qquad
   \left[F_{kl}\right]^L(x_0,\Vec{x})
   \xrightarrow{\mathcal{P}}+\left[F_{kl}\right]^L(x_0,-\Vec{x}).
\label{eq:(1.17)}
\end{equation}

\section{Lattice SUSY current and a renormalized SUSY WT relation}
\label{sec:2}
In this section, we recall how the proposal of~Ref.~\cite{Curci:1986sm} that
SUSY in the lattice formulation of 4D $\mathcal{N}=1$ SYM is restored with an
appropriate tuning of the gluino mass~$M$ is understood in terms of the SUSY WT
relation~\cite{Curci:1986sm,Taniguchi:1999fc,Farchioni:2001wx,Suzuki:2012pc}.
See also Refs.~\cite{Farchioni:2000kb,Farchioni:2001yr}. This discussion
provides the definition of a lattice SUSY current that is conserved in the
continuum limit and our basic SUSY WT relation that will be fully utilized
in the construction of the energy-momentum tensor in~Sec.~\ref{sec:3}.

First, as a lattice counterpart of the SUSY transformation in the
continuum~\eqref{eq:(1.6)}, we define~\cite{Taniguchi:1999fc}
\begin{align}
   \Bar{\delta}_\xi U_\mu(x)&\equiv iag\frac{1}{2}
   \Bar{\xi}\gamma_\mu\left[
   \psi(x)U_\mu(x)
   +U_\mu(x)\psi(x+a\hat\mu)
   \right],
\notag\\
   \Bar{\delta}_\xi U_\mu^\dagger(x)&\equiv-iag\frac{1}{2}
   \Bar{\xi}\gamma_\mu\left[
   U_\mu^\dagger(x)\psi(x)
   +\psi(x+a\hat\mu)U_\mu^\dagger(x)
   \right],
\notag\\
   \Bar{\delta}_\xi\psi(x)&\equiv-\frac{1}{2}\sigma_{\mu\nu}\xi
   \left[F_{\mu\nu}\right]^L(x),\qquad
   \Bar{\delta}_\xi\Bar{\psi}(x)
   =\frac{1}{2}\Bar{\xi}\sigma_{\mu\nu}\left[F_{\mu\nu}\right]^L(x).
\label{eq:(2.1)}
\end{align}
To derive a WT relation, we need also the localized transformation and,
corresponding to the above, we define
\begin{align}
   \delta_\xi U_\mu(x)&\equiv iag\frac{1}{2}
   \left[
   \Bar{\xi}(x)\gamma_\mu\psi(x)U_\mu(x)
   +\Bar{\xi}(x+a\hat\mu)\gamma_\mu U_\mu(x)\psi(x+a\hat\mu)
   \right],
\notag\\
   \delta_\xi U_\mu^\dagger(x)&\equiv-iag\frac{1}{2}
   \left[
   \Bar{\xi}(x)\gamma_\mu U_\mu^\dagger(x)\psi(x)
   +\Bar{\xi}(x+a\hat\mu)\gamma_\mu\psi(x+a\hat\mu)U_\mu^\dagger(x)
   \right],
\notag\\
   \delta_\xi\psi(x)&\equiv-\frac{1}{2}\sigma_{\mu\nu}\xi(x)
   \left[F_{\mu\nu}\right]^L(x),\qquad
   \delta_\xi\Bar{\psi}(x)
   =\frac{1}{2}\Bar{\xi}(x)\sigma_{\mu\nu}\left[F_{\mu\nu}\right]^L(x),
\label{eq:(2.2)}
\end{align}
where the Grassmann-odd spinor parameter~$\xi(x)$ obeys
$\Bar{\xi}(x)=\xi^T(x)(-C^{-1})$.

Now the lattice action~\eqref{eq:(1.10)} is not invariant
under~$\Bar{\delta}_\xi$~\eqref{eq:(2.1)}, because, first of all, the
gluino mass term~\eqref{eq:(1.12)} explicitly breaks SUSY and, various $O(a)$
lattice artifacts in the lattice action also break SUSY. Since the localized
SUSY transformation~$\delta_\xi$~\eqref{eq:(2.2)} reduces to the global
one~\eqref{eq:(2.1)} for~$\xi(x)\to\xi$, the variation of the lattice
action~\eqref{eq:(1.10)} under~Eq.~\eqref{eq:(2.2)} can be written as
\begin{equation}
   \delta_\xi S=a^4\sum_x\Bar{\xi}(x)
   \left[-\partial_\mu^SS_\mu(x)+M\chi(x)+X_S(x)\right],
\label{eq:(2.3)}
\end{equation}
where combinations $S_\mu(x)$ and~$\chi(x)$ are defined by
\begin{equation}
   S_\mu(x)\equiv-\sigma_{\rho\sigma}\gamma_\mu
   \tr\left\{\psi(x)\left[F_{\rho\sigma}\right]^L(x)\right\},
\label{eq:(2.4)}
\end{equation}
and
\begin{equation}
   \chi(x)\equiv\sigma_{\mu\nu}
   \tr\left\{\psi(x)\left[F_{\mu\nu}\right]^L(x)\right\},
\label{eq:(2.5)}
\end{equation}
respectively. $S_\mu(x)$ in~Eq.~\eqref{eq:(2.4)} is nothing but a lattice
transcription of the SUSY current in the continuum, $\Breve{S}_\mu(x)$
in~Eq.~\eqref{eq:(1.7)}, and $\chi(x)$ in~Eq.~\eqref{eq:(2.5)} is an explicit
SUSY breaking caused by~$S_{\text{mass}}$. Thus, in~Eq.~\eqref{eq:(2.3)},
$X_S(x)$ represents an $O(a)$ SUSY breaking effect attributed to the lattice
artifacts. Thus, considering the infinitesimal variation~\eqref{eq:(2.2)} in
the integration variables of the functional integral, we have an exact identity
that holds for any operator~$\mathcal{O}$,\footnote{Throughout this article, we
always assume that an operator represented by the symbol~$\mathcal{O}$ is gauge
invariant.}
\begin{equation}
   \left\langle\partial_\mu^SS_\mu(x)\mathcal{O}\right\rangle
   =\left\langle\left[M\chi(x)+ X_S(x)\right]\mathcal{O}\right\rangle
   -\left\langle
   \frac{1}{a^4}\frac{\partial}{\partial\Bar{\xi}(x)}\delta_\xi
   \mathcal{O}\right\rangle.
\label{eq:(2.6)}
\end{equation}

Here is a remark on the property of~$X_S(x)$ under the parity: From
Eqs.~\eqref{eq:(2.4)} and~\eqref{eq:(2.5)}, we see that
\begin{equation}
   S_\mu(x_0,\Vec{x})\xrightarrow{\mathcal{P}}
   \begin{cases}
   +i\gamma_0S_0(x_0,-\Vec{x})&\text{for $\mu=0$},\\
   -i\gamma_0S_k(x_0,-\Vec{x})&\text{for $\mu=k$},\\
   \end{cases}
\label{eq:(2.7)}
\end{equation}
and
\begin{equation}
   \chi(x_0,\Vec{x})\xrightarrow{\mathcal{P}}
   i\gamma_0\chi(x_0,-\Vec{x}).
\label{eq:(2.8)}
\end{equation}
These shows that $\partial_\mu^SS_\mu(x)$ possesses the same transformation
property as~$\chi(x)$ under the parity. On the other hand, by defining
$\xi(x_0,\Vec{x})\xrightarrow{\mathcal P}i\gamma_0\xi(x_0,-\Vec{x})$
and~$\Bar{\xi}(x_0,\Vec{x})\xrightarrow{\mathcal P}%
-i\Bar{\xi}(x_0,-\Vec{x})\gamma_0$, we see that $\delta_\xi$
in~Eq.~\eqref{eq:(2.2)} and the parity commute to each other. Then by the
parity invariance of the action~$S$ and Eqs.~\eqref{eq:(2.7)}
and~\eqref{eq:(2.8)}, $X_S(x)$ in~Eq.~\eqref{eq:(2.3)} obeys 
\begin{equation}
   X_S(x_0,\Vec{x})\xrightarrow{\mathcal{P}}i\gamma_0X_S(x_0,-\Vec{x}).
\label{eq:(2.9)}
\end{equation}

Next, we consider the renormalization of the composite operator~$X_S(x)$ (just
like for the chiral symmetry~\cite{Bochicchio:1985xa,Testa:1998ez}). It
can be argued that~\cite{Curci:1986sm,Farchioni:2001wx,Suzuki:2012pc}, by using
the assumed symmetries of~$S$ and resultant properties of~$X_S(x)$ (such
as~Eq.~\eqref{eq:(2.9)}), by writing\footnote{In~Ref.~\cite{Suzuki:2012pc},
this operator renormalization was described by using an operator basis in the
continuum theory. With the present lattice regularization, however, it is more
appropriate to use an operator basis in the lattice theory
as~Eq.~\eqref{eq:(2.10)}, because the renormalization is carried out in terms
of counterterms with lattice
structure~\cite{Reisz:1987da,Reisz:1987pw,Luscher:1988sd}. Also, we adopt a
different choice of basis for the term proportional to the equation of motion
from~Ref.~\cite{Suzuki:2012pc}, as this choice is practically more convenient.
Finally, we make use of a simpler lattice SUSY current than
Ref.~\cite{Taniguchi:1999fc}. Then the symmetric difference~$\partial_\mu^S$
instead of the backward difference~$\partial_\mu^*$ appears for the covariance
under the parity.}
\begin{align}
   X_S(x)&=(1-\mathcal{Z}_S)\partial_\mu^SS_\mu(x)
   -\mathcal{Z}_T\partial_\mu^ST_\mu(x)
   -\frac{1}{a}\mathcal{Z}_\chi\chi(x)
\notag\\
   &\qquad{}
   -\mathcal{Z}_{3F}\tr\left[\psi(x)\Bar{\psi}(x)\psi(x)\right]
\notag\\
   &\qquad{}
   -\mathcal{Z}_{\text{EOM}}
   \sigma_{\mu\nu}\tr\{[F_{\mu\nu}]^L(x)(D+M)\psi(x)\}
\notag\\
   &\qquad{}
   +a\mathcal{E}(x),
\label{eq:(2.10)}
\end{align}
and choosing renormalization constants $\mathcal{Z}_S$, $\mathcal{Z}_T$,
$\mathcal{Z}_\chi$, $\mathcal{Z}_{3F}$ and~$\mathcal{Z}_{\text{EOM}}$
appropriately, the dimension~$11/2$ operator $\mathcal{E}(x)$ can be made at
most logarithmically divergent for~$a\to0$. More precisely, the operator
$\mathcal{E}(x)$ is given by a linear combination of renormalized operators
with logarithmically divergent coefficients. Since the operator~$\mathcal{O}$
in~Eq.~\eqref{eq:(2.6)} is gauge invariant, the mixing of~$X_S(x)$ with gauge
non-invariant operators~\cite{Taniguchi:1999fc} does not occur
in~Eq.~\eqref{eq:(2.10)}. A new lattice operator~$T_\mu(x)$
in~Eq.~\eqref{eq:(2.10)} is defined by
\begin{equation}
   T_\mu(x)\equiv2\gamma_\nu
   \tr\left\{\psi(x)\left[F_{\mu\nu}\right]^L(x)\right\},
\label{eq:(2.11)}
\end{equation}
and, from this definition,
\begin{equation}
   T_\mu(x_0,\Vec{x})\xrightarrow{\mathcal{P}}
   \begin{cases}
   +i\gamma_0T_0(x_0,-\Vec{x})&\text{for $\mu=0$},\\
   -i\gamma_0T_k(x_0,-\Vec{x})&\text{for $\mu=k$}.\\
   \end{cases}
\label{eq:(2.12)}
\end{equation}

From Eqs.~\eqref{eq:(2.6)} and~\eqref{eq:(2.10)}, we have
\begin{align}
   &\left\langle
   \partial_\mu^S
   \left[\mathcal{Z}_SS_\mu(x)+\mathcal{Z}_TT_\mu(x)\right]
   \mathcal{O}\right\rangle
\notag\\
   &=\left(M-\frac{1}{a}\mathcal{Z}_\chi\right)
   \left\langle
   \chi(x)\mathcal{O}\right\rangle
   -\mathcal{Z}_{3F}
   \left\langle\tr\left[\psi(x)\Bar{\psi}(x)\psi(x)\right]
   \mathcal{O}\right\rangle
\notag\\
   &\qquad{}
   -\left\langle
   \frac{1}{a^4}\frac{\partial}{\partial\Bar{\xi}(x)}\delta_\xi
   \mathcal{O}\right\rangle
   -\mathcal{Z}_{\text{EOM}}\left\langle
   \sigma_{\mu\nu}\tr\{[F_{\mu\nu}]^L(x)(D+M)\psi(x)\}
   \mathcal{O}\right\rangle
\notag\\
   &\qquad\qquad{}+\left\langle a\mathcal{E}(x)
   \mathcal{O}\right\rangle.
\label{eq:(2.13)}
\end{align}
In this expression, we first note
\begin{equation}
   \left\langle
   \sigma_{\mu\nu}\tr\{[F_{\mu\nu}]^L(x)(D+M)\psi(x)\}
   \mathcal{O}\right\rangle
   =\left\langle
   \frac{1}{a^4}\frac{\partial}{\partial\Bar{\xi}(x)}\delta_{F\xi}
   \mathcal{O}\right\rangle,
\label{eq:(2.14)}
\end{equation}
where the localized transformation~$\delta_{F\xi}$ is a super-like
transformation that acts non-trivially only on the gluino:
\begin{equation}
   \delta_{F\xi}U_\mu(x)=0,\qquad
   \delta_{F\xi}\psi(x)=\delta_\xi\psi(x),\qquad
   \delta_{F\xi}\Bar{\psi}(x)=\delta_\xi\Bar{\psi}(x),
\label{eq:(2.15)}
\end{equation}
because the left-hand side of~Eq.~\eqref{eq:(2.14)} is proportional to the
equation of motion of the gluino (i.e., this is the Schwinger--Dyson (SD)
equation). Therefore, we have
\begin{align}
   &\left\langle
   \partial_\mu^S
   \left[\mathcal{Z}_SS_\mu(x)+\mathcal{Z}_TT_\mu(x)\right]
   \mathcal{O}\right\rangle
\notag\\
   &\qquad=\left(M-\frac{1}{a}\mathcal{Z}_\chi\right)
   \left\langle
   \chi(x)\mathcal{O}\right\rangle
   -\mathcal{Z}_{3F}
   \left\langle\tr\left[\psi(x)\Bar{\psi}(x)\psi(x)\right]
   \mathcal{O}\right\rangle
\notag\\
   &\qquad\qquad{}
   +\left\langle
   \left[-\frac{1}{a^4}\frac{\partial}{\partial\Bar{\xi}(x)}\Delta_\xi
   +a\mathcal{E}(x)\right]
   \mathcal{O}\right\rangle,
\label{eq:(2.16)}
\end{align}
where $\Delta_\xi$ is a \emph{modified\/} localized super transformation,
\begin{equation}
   \Delta_\xi\equiv
   \delta_\xi+\mathcal{Z}_{\text{EOM}}\delta_{F\xi}.
\label{eq:(2.17)}
\end{equation}

Let us consider a special case of~Eq.~\eqref{eq:(2.16)} that $\mathcal{O}$
is a collection of renormalized local operators and the point~$x$ stays away
from the support of~$\mathcal{O}$ by a finite physical distance. In what
follows, we express this situation by
\begin{equation}
   x\leftrightsquigarrow\supp(\mathcal{O}).
\label{eq:(2.18)}
\end{equation}
Under this situation, the $\Bar{\xi}(x)$-derivative in the last line
of~Eq.~\eqref{eq:(2.16)} identically vanishes.\footnote{Our argument below
works when the operator~$\mathcal{O}$ is ultra-local, i.e., its support is
a strictly-finite region on the lattice. With the use of the overlap lattice
Dirac operator~$D$~\cite{Neuberger:1997fp,Neuberger:1998wv}, however,
one might be interested in an operator~$\mathcal{O}$ that is not ultra-local
but exponentially local~\cite{Hernandez:1998et}. Although we do not give any
analysis for such a case, on physical grounds, we believe that our conclusions
will not change.} Also the last term of~Eq.~\eqref{eq:(2.16)} vanishes in the
continuum limit, because for~$x\leftrightsquigarrow\supp(\mathcal{O})$ the
dimension~$11/2$ operator~$\mathcal{E}(x)$ does not produce any $O(1/a)$
divergence that can compensate the overall factor of~$a$. We thus have
\begin{align}
   &\left\langle
   \partial_\mu^S
   \left[\mathcal{Z}_SS_\mu(x)+\mathcal{Z}_TT_\mu(x)\right]
   \mathcal{O}\right\rangle
\notag\\
   &\qquad{}\xrightarrow{a\to0}
   \left(M-\frac{1}{a}\mathcal{Z}_\chi\right)
   \left\langle
   \chi(x)\mathcal{O}\right\rangle
   -\mathcal{Z}_{3F}
   \left\langle\tr\left[\psi(x)\Bar{\psi}(x)\psi(x)\right]
   \mathcal{O}\right\rangle,
\notag\\
   &\qquad\qquad\text{for $x\leftrightsquigarrow\supp(\mathcal{O})$}.
\label{eq:(2.19)}
\end{align}
This can be regarded as a would-be conservation law of a lattice SUSY current.
If this relation with~$\mathcal{Z}_{\text{3F}}\neq0$ held, then the last term,
that is cubic in the gluino field, would give rise to an ``exotic'' SUSY
anomaly. This is not what we expect and we thus assume its absence:
\begin{equation}
   \mathcal{Z}_{3F}=0.
\label{eq:(2.20)}
\end{equation}
As shown in~Ref.~\cite{Suzuki:2012pc} by utilizing the generalized BRS
transformation, this is actually the case at least to all orders in the
perturbation theory. Accepting Eq.~\eqref{eq:(2.20)}, we then have
\begin{align}
   &\left\langle
   \partial_\mu^S
   \left[\mathcal{Z}_SS_\mu(x)+\mathcal{Z}_TT_\mu(x)\right]
   \mathcal{O}\right\rangle
   \xrightarrow{a\to0}
   \left(M-\frac{1}{a}\mathcal{Z}_\chi\right)
   \left\langle
   \chi(x)\mathcal{O}\right\rangle,
\notag\\
   &\qquad{}\text{for $x\leftrightsquigarrow\supp(\mathcal{O})$}.
\label{eq:(2.21)}
\end{align}
The combination $M-(1/a)\mathcal{Z}_\chi$ in the right-hand side corresponds to
an additive renormalization of the gluino mass. So we tune the bare mass
parameter~$M$~\cite{Curci:1986sm} so that
\begin{equation}
   M-\frac{1}{a}\mathcal{Z}_\chi=0.
\label{eq:(2.22)}
\end{equation}
In actual numerical simulations, this tuning should be carried out for each
value of the lattice spacing~$a$. In this way, we obtain the conservation
law in the continuum limit:
\begin{equation}
   \left\langle
   \partial_\mu^S
   \left[\mathcal{Z}_SS_\mu(x)+\mathcal{Z}_TT_\mu(x)\right]
   \mathcal{O}\right\rangle
   \xrightarrow{a\to0}0,
   \qquad\text{for $x\leftrightsquigarrow\supp(\mathcal{O})$},
\label{eq:(2.23)}
\end{equation}
that is a minimal requirement for the restoration of SUSY in the continuum
limit.

We consider the renormalization of the conserved current
$\mathcal{Z}_SS_\mu(x)+\mathcal{Z}_TT_\mu(x)$. Since this is a dimension~$7/2$
gauge invariant vectorial operator, it can mix only with the operators
$S_\mu(x)$ and~$T_\mu(x)$. However, if the mixing were generic, it would be
inconsistent with the conservation law~\eqref{eq:(2.23)}. That is, the
renormalization must be multiplicative.\footnote{It is very interesting to
confirm this renormalizability by the lattice perturbation theory along the
line of~Ref.~\cite{Taniguchi:1999fc}.} Thus we set
\begin{equation}
   \mathcal{S}_\mu(x)
   \equiv\mathcal{Z}
   \left[\mathcal{Z}_SS_\mu(x)+\mathcal{Z}_TT_\mu(x)\right],
\label{eq:(2.24)}
\end{equation}
and call this a renormalized lattice SUSY current. With an appropriate
choice of the constant~$\mathcal{Z}$, $\mathcal{S}_\mu(x)$ should have finite
correlation functions with any renormalized operator~$\mathcal{O}$, as
far as~$x\leftrightsquigarrow\supp(\mathcal{O})$. The renormalization constants
$\mathcal{Z}_S$ and~$\mathcal{Z}_T$ in~Eqs.~\eqref{eq:(2.10)}
and~\eqref{eq:(2.24)} emerge from the power-divergence subtraction and it can
be argued that such constants are independent of the renormalization
scale~$\mu^2$ and finite as~$a\to0$~\cite{Testa:1998ez}. On the other hand,
by the dimensional reason the constant~$\mathcal{Z}$ in~Eq.~\eqref{eq:(2.24)}
is at most logarithmically divergent. In terms of the renormalized SUSY
current~\eqref{eq:(2.24)}, the SUSY WT relation~\eqref{eq:(2.16)} under the
assumptions~\eqref{eq:(2.20)} and~\eqref{eq:(2.22)} reads
\begin{equation}
   \left\langle
   \partial_\mu^S\mathcal{S}_\mu(x)
   \mathcal{O}\right\rangle
   =\left\langle
   \mathcal{Z}
   \left[-\frac{1}{a^4}\frac{\partial}{\partial\Bar{\xi}(x)}\Delta_\xi
   +a\mathcal{E}(x)\right]
   \mathcal{O}\right\rangle.
\label{eq:(2.25)}
\end{equation}

Finally, we argue that for any renormalized local operator~$\mathcal{O}$, the
combination
\begin{equation}
   \mathcal{Z}
   \left[-\frac{1}{a^4}\frac{\partial}{\partial\Bar{\xi}(x)}\Delta_\xi
   +a\mathcal{E}(x)\right]
   \mathcal{O},
\label{eq:(2.26)}
\end{equation}
when $x\in\supp(\mathcal{O})$, produces a renormalized local operator in the
support of~$\mathcal{O}$. In~Eq.~\eqref{eq:(2.26)}, something non-trivial in
the continuum limit can occur only when $x\in\supp(\mathcal{O})$, because
otherwise the $\Bar{\xi}(x)$-derivative vanishes and the dimension~$11/2$
operator~$\mathcal{E}(x)$ does not produce any $O(1/a)$ divergence that can
compensate the overall factor of~$a$. To see what happens when
$x\in\supp(\mathcal{O})$, we sum Eq.~\eqref{eq:(2.25)} over the point~$x$
within a finite region~$\mathcal{D}_{\mathcal{O}}$ that contains the
operator~$\mathcal{O}$. Then, since
$\sum_{x\in\mathcal{D}_{\mathcal{O}}}\partial_\mu^S\mathcal{S}_\mu(x)$ in the
left-hand side becomes a collection of (mutually non-overlapping) renormalized
operators that have no overlap with~$\mathcal{O}$, the left-hand side remains
finite as $a\to0$. This shows that the sum~$\sum_{x\in\mathcal{D}_{\mathcal{O}}}$
of the right-hand side of~Eq.~\eqref{eq:(2.25)} is finite too and we see that
the combination~\eqref{eq:(2.26)} does not produce any ultraviolet divergence
even for $x\in\supp(\mathcal{O})$. We note that such a finiteness will be
indispensable for the existence of a renormalized SYM, because for that there
should exist a renormalized SUSY current that generates a finite super
transformation on renormalized fields. Since the above analysis shows that
$\mathcal{S}_\mu(x)$ is the unique conserved SUSY current in the present
framework, $\mathcal{S}_\mu(x)$ should generate such a finite super
transformation. This argument on the basis of the existence of a renormalized
SYM also supports the finiteness of the combination~\eqref{eq:(2.26)}.

\section{A lattice energy-momentum tensor and its conservation law}
\label{sec:3}
We now define an energy-momentum tensor~$\mathcal{T}_{\mu\nu}(x)$ on the lattice
through the relation quite analogous to~Eq.~\eqref{eq:(1.9)}. That is,
\begin{align}
   \mathcal{Z}\Bar{\Delta}_\xi\mathcal{S}_\mu(x)
   &\equiv
   2\gamma_\nu\xi\bigl\{
   \mathcal{T}_{\mu\nu}(x)+c\delta_{\mu\nu}
   \tr\left[\Bar{\psi}(x)(D+M)\psi(x)\right]
\notag\\
   &\qquad\qquad{}
   +(\text{terms anti-symmetric in $\mu$ and~$\nu$})\bigr\}
\notag\\
   &\qquad{}
   +(\text{terms proportional to $\gamma_5\gamma_\nu\xi$, $\xi$,
   $\gamma_5\xi$, $\sigma_{\nu\rho}\xi$}),
\label{eq:(3.1)}
\end{align}
where $c$~is a constant and $\Bar{\Delta}_\xi$ is a ``global version'' of the
localized super transformation~\eqref{eq:(2.17)}, i.e.,
\begin{equation}
   \Bar{\Delta}_\xi\equiv\Bar{\delta}_\xi
   +\mathcal{Z}_{\text{EOM}}\Bar{\delta}_{F\xi},
\label{eq:(3.2)}
\end{equation}
and $\Bar{\delta}_{F\xi}$ is obtained by setting~$\xi(x)\to\xi$
in~Eq.~\eqref{eq:(2.15)}. One can invert Eq.~\eqref{eq:(3.1)} with respect
to~$\mathcal{T}_{\mu\nu}(x)$: We introduce (here and in what follows, indices
$\alpha$, $\beta$, \dots, which run over $1$, $2$, $3$, $4$, denote the spinor
indices),
\begin{align}
   \varTheta_{\mu\nu}(x)
   &\equiv\frac{1}{8}(\gamma_\nu)_{\beta\alpha}
   \frac{\partial}{\partial\xi_\beta}
   \left[\mathcal{Z}\Bar{\Delta}_\xi\mathcal{S}_\mu(x)\right]_\alpha
\notag\\
   &=\mathcal{Z}^2\mathcal{Z}_S
   \frac{1}{8}(\gamma_\nu)_{\beta\alpha}
   \frac{\partial}{\partial\xi_\beta}
   \left\{\left(\Bar{\delta}_\xi
   +\mathcal{Z}_{\text{EOM}}\Bar{\delta}_{F\xi}\right)\left[
   S_\mu(x)+\frac{\mathcal{Z}_T}{\mathcal{Z}_S}T_\mu(x)
   \right]\right\}_\alpha,
\label{eq:(3.3)}
\end{align}
where in the second equality we have substituted Eqs.~\eqref{eq:(3.2)}
and~\eqref{eq:(2.24)}, then
\begin{equation}
   \mathcal{T}_{\mu\nu}(x)
   =\frac{1}{2}\left[\varTheta_{\mu\nu}(x)+\varTheta_{\nu\mu}(x)\right]
   -c\delta_{\mu\nu}
   \tr\left[\Bar{\psi}(x)(D+M)\psi(x)\right].
\label{eq:(3.4)}
\end{equation}

From relations obtained so far, one can confirm that $\varTheta_{\mu\nu}(x)$
in~Eq.~\eqref{eq:(3.3)} behaves as a second-rank tensor under the
parity:\footnote{To see this, it is convenient to note that $\Bar{\Delta}_\xi$
and the parity commute to each other, if one assigns
$\xi\xrightarrow{\mathcal{P}}i\gamma_0\xi$ and
$\Bar{\xi}\xrightarrow{\mathcal{P}}-i\Bar{\xi}\gamma_0$.}
\begin{align}
   \varTheta_{\mu\nu}(x_0,\Vec{x})&\xrightarrow{\mathcal{P}}
   \begin{cases}
   +\varTheta_{00}(x_0,-\Vec{x})&\text{for $\mu=\nu=0$},\\
   -\varTheta_{k0}(x_0,-\Vec{x})&\text{for $\mu=k$ and $\nu=0$},\\
   -\varTheta_{0k}(x_0,-\Vec{x})&\text{for $\mu=0$ and $\nu=k$},\\
   +\varTheta_{kl}(x_0,-\Vec{x})&\text{for $\mu=k$ and $\nu=l$}.
   \end{cases}
\label{eq:(3.5)}
\end{align}

The defining relation~\eqref{eq:(3.1)} is quite analogous to the classical
relation~\eqref{eq:(1.9)}. Yet they differ in that the super
transformation~$\Bar{\delta}_\xi$ is replaced by the renormalized modified
super transformation~$\mathcal{Z}\Bar{\Delta}_\xi$ and the Dirac operator is
shifted by the gluino mass~$M$; also the coefficient of the term that is
proportional to the fermion action is changed from~$3/4$ to~$c$.

Quite interestingly, we can show that the energy-momentum tensor defined
by~Eq.~\eqref{eq:(3.4)} is conserved in the quantum continuum limit. That is
\begin{equation}
   \left\langle\partial_\mu^S\mathcal{T}_{\mu\nu}(x)\mathcal{O}\right\rangle
   \xrightarrow{a\to0}0,
   \qquad\text{for $x\leftrightsquigarrow\supp(\mathcal{O})$},
\label{eq:(3.6)}
\end{equation}
for any renormalized local operator~$\mathcal{O}$. Note that this conservation
law holds for any value of~$c$ in~Eq.~\eqref{eq:(3.4)} because the term
$\tr[\Bar{\psi}(x)(D+M)\psi(x)]$ is proportional to the equation of motion and
it has no correlation with~$\mathcal{O}$ when $x$ is not in the support
of~$\mathcal{O}$. When $x\notin\supp(\mathcal{O})$, one can replace it by
\begin{equation}
   \left\langle\tr\left[\Bar{\psi}(x)(D+M)\psi(x)\right]\right\rangle
   =-2(N_c^2-1)a^{-4}.
\label{eq:(3.7)}
\end{equation}
Since this is independent of~$x$, the last term in~Eq.~\eqref{eq:(3.4)} does not
affect the conservation law. The most general structure of the energy-momentum
tensor (that is symmetric in its indices) in the lattice formulation of the
Yang--Mills theory coupled to fermions has been given
in~Ref.~\cite{Caracciolo:1989pt}; see also
Refs.~\cite{Caracciolo:1989bu,Caracciolo:1991vc,Caracciolo:1991cp} for
related analyses. According to~Ref.~\cite{Caracciolo:1989pt}, a symmetric
energy-momentum tensor that satisfies the conservation law in the continuum
limit~\eqref{eq:(3.6)}, if it exists, is essentially unique. Only ambiguities
are the overall normalization ($\mathcal{Z}^2\mathcal{Z}_S$
in~Eq.~\eqref{eq:(3.3)} in our construction) and a proportionality constant to
the fermion Lagrangian (i.e., the constant $c$ in~Eq.~\eqref{eq:(3.4)}). The
bottom line is that our definition~\eqref{eq:(3.4)}, that is suggested from the
structure of the FZ supermultiplet, gives rise to a very explicit form of this
unique conserved symmetric energy-momentum tensor.

To show the conservation law~\eqref{eq:(3.6)}, we first show the conservation
law of~$\varTheta_{\mu\nu}(x)$ in~Eq.~\eqref{eq:(3.3)} that is not necessary
symmetric under~$\mu\leftrightarrow\nu$. Then we show the anti-symmetric
part of~$\varTheta_{\mu\nu}(x)$ also is conserved in the continuum limit,
implying~Eq.~\eqref{eq:(3.6)}.

To show the conservation law of~$\varTheta_{\mu\nu}(x)$, we set
$\mathcal{O}\to\partial_\nu^S\mathcal{S}_\nu(y)\mathcal{O}$ in the WT
relation~\eqref{eq:(2.25)}. This yields
\begin{align}
   &\left\langle
   \left[\partial_\mu^S\mathcal{S}_\mu(x)\right]_\alpha
   \left[\partial_\nu^S\mathcal{S}_\nu(y)\right]_\beta
   \mathcal{O}\right\rangle
\notag\\
   &=\left\langle
   \mathcal{Z}\left[
   -\frac{1}{a^4}\frac{\partial}{\partial\Bar{\xi}(x)}\Delta_\xi
   +a\mathcal{E}(x)\right]_\alpha
   \left[\partial_\nu^S\mathcal{S}_\nu(y)\right]_\beta
   \mathcal{O}\right\rangle
\notag\\
   &=-\left\langle
   \mathcal{Z}
   \left[
   -\frac{1}{a^4}\frac{\partial}{\partial\Bar{\xi}(y)}\Delta_\xi
   +a\mathcal{E}(y)\right]_\beta
   \left[\partial_\mu^S\mathcal{S}_\mu(x)\right]_\alpha
   \mathcal{O}\right\rangle
\notag\\
   &=\left\langle
   \frac{1}{a^4}\frac{\partial}{\partial\Bar{\xi}_\beta(y)}
   \left[\mathcal{Z}\Delta_\xi\partial_\mu^S\mathcal{S}_\mu(x)\right]_\alpha
   \mathcal{O}\right\rangle
\notag\\
   &\qquad{}+\left\langle
   \left[\partial_\mu^S\mathcal{S}_\mu(x)\right]_\alpha
   \mathcal{Z}\left[
   -\frac{1}{a^4}\frac{\partial}{\partial\Bar{\xi}(y)}\Delta_\xi
   +a\mathcal{E}(y)\right]_\beta
   \mathcal{O}\right\rangle.
\label{eq:(3.8)}
\end{align}
The second equality holds because the first expression is anti-symmetric under
the exchange, $x\leftrightarrow y$, $\mu\leftrightarrow\nu$
and~$\alpha\leftrightarrow\beta$. Using Eq.~\eqref{eq:(2.25)} once again in
the last line, we have the identity
\begin{align}
   &\left\langle
   \frac{1}{a^4}\frac{\partial}{\partial\Bar{\xi}_\beta(y)}
   \left[\mathcal{Z}\Delta_\xi\partial_\mu^S\mathcal{S}_\mu(x)\right]_\alpha
   \mathcal{O}\right\rangle
\notag\\
   &=\left\langle
   \mathcal{Z}\left[
   -\frac{1}{a^4}\frac{\partial}{\partial\Bar{\xi}(x)}
   \Delta_\xi+a\mathcal{E}(x)\right]_\alpha
   \left[\partial_\nu^S\mathcal{S}_\nu(y)\right]_\beta
   \mathcal{O}\right\rangle
\notag\\
   &\qquad{}-\left\langle
   \mathcal{Z}\left[
   -\frac{1}{a^4}\frac{\partial}{\partial\Bar{\xi}(x)}
   \Delta_\xi+a\mathcal{E}(x)\right]_\alpha
   \mathcal{Z}\left[
   -\frac{1}{a^4}\frac{\partial}{\partial\Bar{\xi}(y)}
   \Delta_\xi+a\mathcal{E}(y)\right]_\beta
   \mathcal{O}\right\rangle.
\label{eq:(3.9)}
\end{align}
We then sum this relation over~$y$ within a finite region~$\mathcal{D}_x$
containing the operator~$\partial_\mu^S\mathcal{S}_\mu(x)$. Noting the
identity between the local and global transformations,
\begin{equation}
   \sum_{y\in\mathcal{D}_x}\frac{\partial}{\partial\Bar{\xi}_\beta(y)}\Delta_\xi
   \partial_\mu^S\mathcal{S}_\mu(x)
   =\frac{\partial}{\partial\Bar{\xi}_\beta}\Bar{\Delta}_\xi
   \partial_\mu^S\mathcal{S}_\mu(x),
\label{eq:(3.10))}
\end{equation}
we have
\begin{align}
   &\left\langle
   \frac{\partial}{\partial\Bar{\xi}_\beta}
   \left[\mathcal{Z}\Bar{\Delta}_\xi
   \partial_\mu^S\mathcal{S}_\mu(x)\right]_\alpha
   \mathcal{O}\right\rangle
\notag\\
   &=\left\langle
   \mathcal{Z}\left[
   -\frac{1}{a^4}\frac{\partial}{\partial\Bar{\xi}(x)}
   \Delta_\xi+a\mathcal{E}(x)\right]_\alpha
   a^4\sum_{y\in\mathcal{D}_x}\left[\partial_\nu^S\mathcal{S}_\nu(y)\right]_\beta
   \mathcal{O}\right\rangle
\notag\\
   &\qquad{}-\left\langle
   \mathcal{Z}\left[
   -\frac{1}{a^4}\frac{\partial}{\partial\Bar{\xi}(x)}\Delta_\xi
   +a\mathcal{E}(x)\right]_\alpha
   a^4\sum_{y\in\mathcal{D}_x}\mathcal{Z}\left[
   -\frac{1}{a^4}\frac{\partial}{\partial\Bar{\xi}(y)}\Delta_\xi
   +a\mathcal{E}(y)\right]_\beta
   \mathcal{O}\right\rangle.
\label{eq:(3.11)}
\end{align}
Since $\Bar{\xi}=\xi^T(-C^{-1})$ implies
\begin{equation}
   (\gamma_\nu)_{\beta\alpha}
   \frac{\partial}{\partial\xi_\beta}
   \left[\mathcal{Z}\Bar{\Delta}_\xi
   \partial_\mu^S\mathcal{S}_\mu(x)\right]_\alpha
   =(C^{-1}\gamma_\nu)_{\alpha\beta}
   \frac{\partial}{\partial\Bar{\xi}_\beta}
   \left[\mathcal{Z}\Bar{\Delta}_\xi
   \partial_\mu^S\mathcal{S}_\mu(x)\right]_\alpha,
\label{eq:(3.12)}
\end{equation}
Eqs.~\eqref{eq:(3.3)} and~\eqref{eq:(3.11)} yield,
\begin{align}
   &\left\langle
   \partial_\mu^S\varTheta_{\mu\nu}(x)
   \mathcal{O}\right\rangle
\notag\\
   &=\frac{1}{8}
   (C^{-1}\gamma_\nu)_{\alpha\beta}
   \Biggl\langle
   \mathcal{Z}\left[
   -\frac{1}{a^4}\frac{\partial}{\partial\Bar{\xi}(x)}\Delta_\xi
   +a\mathcal{E}(x)\right]_\alpha
   a^4\sum_{y\in\mathcal{D}_x}
   \left[\partial_\nu^S\mathcal{S}_\nu(y)\right]_\beta
   \mathcal{O}\Biggr\rangle
\notag\\
   &\qquad{}
   -\frac{1}{8}
   (C^{-1}\gamma_\nu)_{\alpha\beta}
   \Biggl\langle
   \mathcal{Z}\left[
   -\frac{1}{a^4}\frac{\partial}{\partial\Bar{\xi}(x)}\Delta_\xi
   +a\mathcal{E}(x)\right]_\alpha
\notag\\
   &\qquad\qquad\qquad\qquad\qquad{}
   \times a^4\sum_{y\in\mathcal{D}_x}\mathcal{Z}\left[
   -\frac{1}{a^4}\frac{\partial}{\partial\Bar{\xi}(y)}
   \Delta_\xi+a\mathcal{E}(y)\right]_\beta
   \mathcal{O}\Biggr\rangle.
\label{eq:(3.13)}
\end{align}
Now, let us suppose that the point~$x$ stays away from the support
of the renormalized operator~$\mathcal{O}$ by a finite physical distance, i.e.,
$x\leftrightsquigarrow\supp(\mathcal{O})$. Let us further assume that the
region~$\mathcal{D}_x$ is chosen so that it does not overlap with the support
of~$\mathcal{O}$. In this situation, since
$\sum_{y\in\mathcal{D}_x}\partial_\nu^S\mathcal{S}_\nu(y)$ does not have any
support at~$x$, Eq.~\eqref{eq:(3.13)} reduces to
\begin{align}
   &\left\langle
   \partial_\mu^S\varTheta_{\mu\nu}(x)
   \mathcal{O}\right\rangle
\notag\\
   &=\frac{1}{8}
   (C^{-1}\gamma_\nu)_{\alpha\beta}
   \Biggl\langle
   \mathcal{Z}\left[
   a\mathcal{E}(x)\right]_\alpha
   a^4\sum_{y\in\mathcal{D}_x}
   \left[\partial_\nu^S\mathcal{S}_\nu(y)\right]_\beta
   \mathcal{O}\Biggr\rangle
\notag\\
   &\qquad{}
   -\frac{1}{8}
   (C^{-1}\gamma_\nu)_{\alpha\beta}
   \Biggl\langle
    a^4\sum_{y\in\mathcal{D}_x}
   \mathcal{Z}\left[
   -\frac{1}{a^4}\frac{\partial}{\partial\Bar{\xi}(x)}\Delta_\xi
   +a\mathcal{E}(x)\right]_\alpha
   \mathcal{Z}\left[a\mathcal{E}(y)\right]_\beta
   \mathcal{O}\Biggr\rangle.
\label{eq:(3.14)}
\end{align}
The first term in the right-hand side is a correlation function of renormalized
operators without any mutual overlap with the overall factor of~$a$. Thus, this
term vanishes in the $a\to0$ limit. In the second term in the right-hand side,
we use the property argued around Eq.~\eqref{eq:(2.26)}. Then it is also a
correlation function of renormalized operators without any mutual overlap with
the overall factor of~$a$. Thus this also vanishes in the $a\to0$ limit. In
this way, we conclude
\begin{equation}
   \left\langle\partial_\mu^S\varTheta_{\mu\nu}(x)\mathcal{O}\right\rangle
   \xrightarrow{a\to0}0,
   \qquad\text{for $x\leftrightsquigarrow\supp(\mathcal{O})$},
\label{eq:(3.15)}
\end{equation}
for any renormalized operator~$\mathcal{O}$.

Next, we consider the anti-symmetric part of~$\varTheta_{\mu\nu}(x)$,
\begin{equation}
   \mathcal{A}_{\mu\nu}(x)
   \equiv\frac{1}{2}
   \left[\varTheta_{\mu\nu}(x)-\varTheta_{\nu\mu}(x)\right],
\label{eq:(3.16)}
\end{equation}
and its renormalization. It must be possible to expand this composite operator
by gauge invariant operators with dimensions less than or equal to~$4$ that
behave in the same way under the hypercubic group as the right-hand side
of~Eq.~\eqref{eq:(3.16)} (for the parity, see~Eq.~\eqref{eq:(3.5)}). By taking
also the constraint~\eqref{eq:(1.5)} into account, the most general possibility
turns to be\footnote{For example, a seemingly-obvious candidate
$\tr[\Bar{\psi}(x)\sigma_{\mu\nu}\psi(x)]$ identically vanishes because
of~Eq.~\eqref{eq:(1.5)}; another candidate
$\tr[\Bar{\psi}(x)\gamma_\mu D_\nu^S\psi(x)]-(\mu\leftrightarrow\nu)$, where
$D_\mu^Sf(x)\equiv(1/2a)[U_\mu(x)f(x+a\Hat{\mu})U_\mu^\dagger(x)
-U_\mu(x-a\Hat{\mu})^\dagger f(x-a\Hat{\mu})U_\mu(x-a\Hat{\mu})]$,
reduces to~Eq.~\eqref{eq:(3.17)}, by the fact that $D=\Slash{D}+O(a)$
and~$\sigma_{\mu\nu}\Slash{D}
=-\epsilon_{\mu\nu\rho\sigma}D_\rho\gamma_\sigma\gamma_5
+\gamma_\mu D_\nu-\gamma_\nu D_\mu$.}
\begin{align}
   &\mathcal{A}_{\mu\nu}(x)
\notag\\
   &=A_1\epsilon_{\mu\nu\rho\sigma}\partial_\rho^S
   \tr\left[\Bar{\psi}(x)\gamma_\sigma\gamma_5\psi(x)\right]
   +A_2\tr\left[\Bar{\psi}(x)\sigma_{\mu\nu}(D+M)\psi(x)\right]
   +a\mathcal{G}_{\mu\nu}(x).
\label{eq:(3.17)}
\end{align}
That is, by choosing the constants $A_1$ and~$A_2$ appropriately, the
dimension~$5$ operator~$\mathcal{G}_{\mu\nu}(x)$ can be made at most
logarithmically divergent. Now, quite fortunately, Eq.~\eqref{eq:(3.17)} shows
that the anti-symmetric part is conserved by itself:
\begin{equation}
   \left\langle\partial_\mu^S\mathcal{A}_{\mu\nu}(x)\mathcal{O}\right\rangle
   \xrightarrow{a\to0}0,
   \qquad\text{for $x\leftrightsquigarrow\supp(\mathcal{O})$}.
\label{eq:(3.18)}
\end{equation}
This is trivially true for the first term of~Eq.~\eqref{eq:(3.17)}. For the
second term, this follows from the SD equation (the equation of motion
of~$\psi(x)$). Finally, the last term of~Eq.~\eqref{eq:(3.17)} vanishes in
the continuum limit. Combined Eqs.~\eqref{eq:(3.18)} and~\eqref{eq:(3.15)},
we conclude that the symmetric part of~$\varTheta_{\mu\nu}(x)$,
$(1/2)[\varTheta_{\mu\nu}(x)+\varTheta_{\nu\mu}(x)]$, also is conserved. This
proves our assertion Eq.~\eqref{eq:(3.6)}, because the symmetric part
of~$\varTheta_{\mu\nu}(x)$ is the first term of~Eq.~\eqref{eq:(3.4)} and the
term with the coefficient~$c$ does not affect the conservation law as we
already explained.

We note that, with our definition of the energy-momentum
tensor~\eqref{eq:(3.4)}, the trace
anomaly~\cite{Crewther:1972kn,Chanowitz:1972vd,Adler:1976zt,Nielsen:1977sy}
and the gamma-trace anomaly (superconformal
anomaly)~\cite{Abbott:1977in,Curtright:1977cg,Inagaki:1978iu,Majumdar:1980ej,
Nicolai:1980km,Hagiwara:1979pu,Hagiwara:1980ys,Kumar:1982ng,Nakayama:1983qt}
are related by\footnote{Ref.~\cite{Fujikawa:1983az} is a pioneering work
on the trace anomaly in lattice gauge theory.}
\begin{align}
   \left\langle
   \mathcal{T}_{\mu\mu}(x)\mathcal{O}
   \right\rangle
   =
   \frac{1}{8}
   \left\langle   
   \frac{\partial}{\partial\xi_\alpha}
   \left[\mathcal{Z}\Bar{\Delta}_\xi\gamma_\mu\mathcal{S}_\mu(x)\right]_\alpha
   \mathcal{O}
   \right\rangle
   +8c(N_c^2-1)a^{-4}
   \left\langle   
   \mathcal{O}
   \right\rangle,
\label{eq:(3.19)}
\end{align}
where we have used Eq.~\eqref{eq:(3.7)} assuming $x\notin\supp(\mathcal{O})$.

Now we discuss how our energy-momentum tensor~\eqref{eq:(3.4)} can be used in
actual non-perturbative Monte Carlo simulations. As Eqs.~\eqref{eq:(3.3)}
and~\eqref{eq:(3.4)} show, the definition of the energy-momentum tensor
contains four combinations of renormalization constants,
$\mathcal{Z}^2\mathcal{Z}_S$, $\mathcal{Z}_{\text{EOM}}$,
$\mathcal{Z}_T/\mathcal{Z}_S$ and~$c$; we have to know these numbers to
construct the energy-momentum tensor. Among these, the
ratio~$\mathcal{Z}_T/\mathcal{Z}_S$ has been measured~\cite{Donini:1997hh,%
Farchioni:2000mp,Farchioni:2001yn,Farchioni:2004fy,Demmouche:2010sf} by
determining the coefficients in the relation ($\mathcal{O}$ is a fermionic
spinorial operator)
\begin{equation}
   \partial_\mu^S\left\langle S_\mu(x)\mathcal{O}\right\rangle
   +\frac{\mathcal{Z}_T}{\mathcal{Z}_S}
   \partial_\mu^S\left\langle T_\mu(x)\mathcal{O}\right\rangle
   =\frac{1}{\mathcal{Z}_S}\left(M-\frac{1}{a}\mathcal{Z}_\chi\right)
   \left\langle\chi(x)\mathcal{O}\right\rangle.
\label{eq:(3.20)}
\end{equation}
This is Eq.~\eqref{eq:(2.21)} up to~$O(a)$ corrections. Thus we already know
that the number~$\mathcal{Z}_T/\mathcal{Z}_S$ can be determined numerically (up
to~$O(a)$ corrections). Next, to determine~$\mathcal{Z}_{\text{EOM}}$
in~Eq.~\eqref{eq:(3.3)}, we may use the conservation law
of~$\varTheta_{\mu\nu}(x)$, Eq.~\eqref{eq:(3.15)}, itself.

The determinations of~$\mathcal{Z}^2\mathcal{Z}_S$ and~$c$ are somewhat
correlated. We first note that, when the energy-momentum
tensor~$\mathcal{T}_{\mu\nu}(x)$~\eqref{eq:(3.4)} is inserted in a
physical amplitude, the last term of~Eq.~\eqref{eq:(3.4)} just gives rise to an
additive constant~\eqref{eq:(3.7)} that is identical for \emph{any\/} physical
amplitudes; the constant~$c$ thus corresponds to a choice of the origin of the
energy. If we consider the difference in expectation values
of~$\mathcal{T}_{\mu\nu}(x)$ in two physical states, therefore, the contribution
of the last term of~Eq.~\eqref{eq:(3.4)} cancels out and it is the same as the
difference in expectation values
of~$(1/2)[\varTheta_{\mu\nu}(x)+\varTheta_{\nu\mu}(x)]$. Noting this fact, one
may determine $\mathcal{Z}^2\mathcal{Z}_S$, the absolute normalization
of~$\varTheta_{\mu\nu}(x)$ in~Eq.~\eqref{eq:(3.3)}, by setting the difference of
expectation values of the ``energy operator''
$-a^3\sum_{\Vec{x}}\varTheta_{00}(x)$ in two different physical states
to a certain prescribed value.

Although the value of~$c$ influences neither on the conservation law nor on the
difference in expectation values of the energy-momentum tensor, there exists a
physically natural choice of~$c$ in the present supersymmetric system. That
is, we may require that the expectation value of the energy density to vanish
\begin{equation}
   \left\langle\mathcal{T}_{00}(x)\right\rangle=0,
\label{eq:(3.21)}
\end{equation}
when \emph{periodic boundary conditions} are imposed on all the fields. This
requirement fixes
\begin{equation}
   c=-\frac{a^4}{2(N_c^2-1)}
   \left\langle\varTheta_{00}(x)\right\rangle_{\text{periodic boundary conditions}},
\label{eq:(3.22)}
\end{equation}
because of~Eq.~\eqref{eq:(3.7)}. Eq.~\eqref{eq:(3.21)} states that the
derivative of the supersymmetric partition function (i.e., the Witten
index~\cite{Witten:1982df}) with respect to the temporal size of the system
vanishes; this holds only if a choice of the origin of the energy is consistent
with the SUSY algebra. In other words, Eq.~\eqref{eq:(3.21)} is required for
the spatial integral~$-a^3\sum_{\Vec{x}}\mathcal{T}_{00}(x)$ to be the energy
operator appearing in the right-hand side of the SUSY algebra. It is thus a
natural requirement from the perspective of SUSY.\footnote{We emphasize that
Eq.~\eqref{eq:(3.21)} should hold even in a theory in which SUSY is
spontaneously broken (recall that the Witten index is independent of the
temporal size of the system even in such a case). This same idea was adopted
to find an expression of the energy density in a lattice formulation of the
two-dimensional (2D) $\mathcal{N}=(2,2)$
SYM~\cite{Kanamori:2007ye,Kanamori:2007yx,Kadoh:2009rw}. For this 2D system,
fortunately, there exist lattice formulations that possess one exact
fermionic symmetry~\cite{Cohen:2003xe,Sugino:2003yb,Sugino:2004qd,%
D'Adda:2005zk} and, employing this symmetry, one can define the energy density
operator that shows the zero-point energy consistent with SUSY. This energy
density operator on the lattice has been used~\cite{Kanamori:2009dk} to measure
the vacuum energy density associated with a possible dynamical SUSY breaking in
the above 2D system.} This completes our non-perturbative construction of a
symmetric energy-momentum tensor. It is conserved in the quantum continuum
limit and, by choosing~$c$ as~Eq.~\eqref{eq:(3.22)}, the origin of the energy
is consistent with SUSY.

A remaining important issue is whether our symmetric energy-momentum
tensor~\eqref{eq:(3.4)} generates, through a renormalized WT relation, a
correctly-normalized translation on renormalized fields. Although we do not
go into this question in the present paper, assuming the existence of a
renormalized translational invariant theory and considering the uniqueness of
the conserved symmetric energy-momentum tensor in the continuum
limit~\cite{Caracciolo:1989pt}, we believe that the answer is affirmative:
Of course, further consideration is required on this point.

\section{Conclusion}

We have presented a non-perturbative construction of a symmetric
energy-momentum tensor in the lattice formulation of 4D $\mathcal{N}=1$ SYM.
Inspired by the relation~\eqref{eq:(1.9)} in the FZ supermultiplet, we defined
the energy-momentum tensor by a renormalized modified super transformation of a
renormalized SUSY current that generates a finite super transformation on
renormalized operators (under the tuning of the gluino mass). Then, it can be
shown that the lattice energy-momentum tensor is conserved in the quantum
continuum limit. The resulting energy-momentum tensor may be used to measure
physical quantities related with the energy-momentum tensor (such as the
viscosity) by numerical simulations.

A question naturally arises is then what will be resulted from the
consideration on the another relation in the FZ multiplet,
Eq.~\eqref{eq:(1.8)}, a relation between the $U(1)_A$ current and the SUSY
current. It is clear that the classical relation~\eqref{eq:(1.8)} as it stands
cannot hold in quantum theory, because the $U(1)_A$ current is not conserved
owing to the axial anomaly, while we are believing that SUSY does not suffer
from any anomaly. Ironically, it turns out that the situation for this
seemingly simple relation~\eqref{eq:(1.8)} is much more complicated than for
(seemingly complicated) Eq.~\eqref{eq:(1.9)}; this is precisely because of the
axial anomaly. We hope to come back this problem in the near future. If the
full structure of the FZ multiplet (with corrections by the anomaly) can be
realized in a well-regularized framework such as the lattice, it will be a
quite useful starting point to understand the so-called anomaly
puzzle~\cite{Clark:1978jx,Grisaru:1978vx,Grisaru:1985yk,Grisaru:1985ik,%
Shifman:1986zi,Ensign:1987wy,ArkaniHamed:1997mj,Huang:2010tn,%
Komargodski:2010rb,Yonekura:2010mc,Yonekura:2012uk}.

\section*{Note added in proof}

Eq.~\eqref{eq:(3.13)} implies the following very suggestive relation: Noting
that $\sum_{y\in\mathcal{D}_x}\partial_\nu^S\mathcal{S}_\nu(y)$ in the first term
of the right-hand side of~Eq.~\eqref{eq:(3.13)} does not have the support
at~$y=x$, using the SUSY WT relation~\eqref{eq:(2.25)} again, we have
\begin{align}
   &a^4\sum_{x\in\mathcal{D}_{\mathcal{O}}}
   \left\langle
   \partial_\mu^S\varTheta_{\mu\nu}(x)
   \mathcal{O}\right\rangle
\notag\\
   &=-\frac{1}{8}
   (C^{-1}\gamma_\nu)_{\alpha\beta}\,
   a^4\sum_{x\in\mathcal{D}_{\mathcal{O}}}
   a^4\sum_{y\in\mathcal{D}_x}
\notag\\
   &\qquad{}\times
   \left\langle
   \left\{
   \mathcal{Z}\left[
   -\frac{1}{a^4}\frac{\partial}{\partial\Bar{\xi}(x)}\Delta_\xi
   +a\mathcal{E}(x)\right]_\alpha,
   \mathcal{Z}\left[
   -\frac{1}{a^4}\frac{\partial}{\partial\Bar{\xi}(y)}
   \Delta_\xi+a\mathcal{E}(y)\right]_\beta\right\}
   \mathcal{O}\right\rangle,
\end{align}
where $\mathcal{D}_{\mathcal{O}}$ is a finite region that contains the
operator~$\mathcal{O}$ entirely. On the other hand, for the anti-symmetric part
of~$\varTheta_{\mu\nu}(x)$, $\mathcal{A}_{\mu\nu}(x)$ in~Eq.~\eqref{eq:(3.16)},
\begin{equation}
   a^4\sum_{x\in\mathcal{D}_{\mathcal{O}}}
   \left\langle
   \partial_\mu^S\mathcal{A}_{\mu\nu}(x)
   \mathcal{O}\right\rangle\xrightarrow{a\to0}0,
\end{equation}
because Eq.~\eqref{eq:(3.17)} shows that
$\langle\mathcal{A}_{\mu\nu}(x)\mathcal{O}\rangle$ is proportional to the
delta function~$\delta^4(x-z)$ in the continuum limit, where $z$ is any point
in the support of~$\mathcal{O}$. Similarly, for the last term
of~Eq.~\eqref{eq:(3.4)}, we have
\begin{equation}
   a^4\sum_{x\in\mathcal{D}_{\mathcal{O}}}
   \left\langle
   \partial_\mu^S\delta_{\mu\nu}
   \tr\left[\Bar{\psi}(x)(D+M)\psi(x)\right]
   \mathcal{O}\right\rangle=0.
\end{equation}
Thus, combining the above three relations,
\begin{align}
   &a^4\sum_{x\in\mathcal{D}_{\mathcal{O}}}
   \left\langle
   \partial_\mu^S\mathcal{T}_{\mu\nu}(x)
   \mathcal{O}\right\rangle
\notag\\
   &\xrightarrow{a\to0}-\frac{1}{8}
   (C^{-1}\gamma_\nu)_{\alpha\beta}\,
   a^4\sum_{x\in\mathcal{D}_{\mathcal{O}}}
   a^4\sum_{y\in\mathcal{D}_x}
\notag\\
   &\qquad{}\times
   \left\langle
   \left\{
   \mathcal{Z}\left[
   -\frac{1}{a^4}\frac{\partial}{\partial\Bar{\xi}(x)}\Delta_\xi
   +a\mathcal{E}(x)\right]_\alpha,
   \mathcal{Z}\left[
   -\frac{1}{a^4}\frac{\partial}{\partial\Bar{\xi}(y)}
   \Delta_\xi+a\mathcal{E}(y)\right]_\beta\right\}
   \mathcal{O}\right\rangle.
\end{align}
This may be regarded as the SUSY algebra in the present lattice framework if
we assume that the lattice energy-momentum tensor~$\mathcal{T}_{\mu\nu}$
generates the translation (on gauge invariant operators~$\mathcal{O}$) as
\begin{align}
   a^4\sum_{x\in\mathcal{D}_{\mathcal{O}}}
   \left\langle
   \partial_\mu^S\mathcal{T}_{\mu\nu}(x)
   \mathcal{O}\right\rangle
   \xrightarrow{a\to0}
   -\left\langle\partial_\nu\mathcal{O}\right\rangle.
\end{align}

\section*{Remark in proof}
A quite similar but somewhat different definition of a lattice energy-momentum
tensor in 4D $\mathcal{N}=1$ SYM, which is superior in several aspects compared
with the one in the present note, has been given in~Ref.~\cite{Suzuki:2012wx}.

\section*{Acknowledgements}

I would like to thank Michael G.~Endres, Kazuo Fujikawa, Yoshio Kikukawa,
Martin L\"uscher, Yusuke Taniguchi and Naoto Yokoi for enlightening
discussions. Some part of this work was carried out during the YITP workshop on
``Field Theory and String Theory'' (YITP-W-12-05) and I would like to thank the
Yukawa Institute for Theoretical Physics at Kyoto University for the
hospitality. This work is supported in part by a Grant-in-Aid for Scientific
Research, nos.~22340069 and~23540330.



%% References


\bibliographystyle{elsarticle-num}
\bibliography{<your-bib-database>}



\end{document}